%% file: martini_nuphys16.tex
\documentclass[12pt]{article}
\usepackage{graphicx}
\newcommand\pubnumber{NuPhys2016-Martini}
\newcommand\pubdate{\today}

\def\esnt{ESNT, CEA, IRFU, Service de Physique Nucl\'eaire, Universit\'e de Paris-Saclay, F-91191 Gif-sur-Yvette, France}
\def\support{\footnote{Work supported by the ``Espace de Structure et
    de r\'eactions Nucl\'eaire Th\'eorique'' (ESNT) %, \url{http://esnt.cea.fr} )
    at CEA.}}

\def\Title#1{\begin{center} {\Large #1 } \end{center}}
\def\Author#1{\begin{center}{ \sc #1} \end{center}}
\def\Address#1{\begin{center}{ \it #1} \end{center}}

\newcommand\pubblock{\rightline{\begin{tabular}{l} \pubnumber\\
         \pubdate  \end{tabular}}}
\newenvironment{Abstract}{\begin{quotation}  }{\end{quotation}}
\newenvironment{Presented}{\begin{quotation} \begin{center} 
             PRESENTED AT\end{center}\bigskip 
      \begin{center}\begin{large}}{\end{large}\end{center} \end{quotation}}
\def\Acknowledgements{\bigskip  \bigskip \begin{center} \begin{large}
             \bf ACKNOWLEDGEMENTS \end{large}\end{center}}
\input econfmacros.tex
\begin{document}
\begin{titlepage}
\pubblock

\vfill
\Title{Status and challenges of neutrino cross sections}
\vfill
\Author{ Marco Martini\support}
\Address{\esnt}
\vfill
\begin{Abstract}
Neutrino oscillations physics entered in the precision era. 
In this context accelerator-based neutrino experiments 
need a reduction of systematic errors to the level of a few percent.
Today one of the most important sources of systematic errors are the neutrino-nucleus cross sections.
The status of our knowledge of these cross sections in the different open channels in the few-GeV region, i.e. the quasielastic, the pion production and the multinucleon emission, is reviewed.
Special emphasis is devoted to the multinucleon emission channel, which attracted a lot of attention in the last few years.
It is crucial to properly reconstruct the neutrino energy which enters the expression of the oscillation probability. 
This channel was not included in the generators used for the analyses of the neutrino cross sections and oscillations experiments. 
\end{Abstract}
\vfill
\begin{Presented}

NuPhys2016, Prospects in Neutrino Physics

Barbican Centre, London, UK,  December 12--14, 2016

\end{Presented}
\vfill
\end{titlepage}
\def\thefootnote{\fnsymbol{footnote}}
\setcounter{footnote}{0}

\section{Introduction}
Neutrino physics has undergone a spectacular development in the last decades, following 
the discovery of neutrino oscillations and nowadays is entered in the precision era which needs a reduction of systematic errors to the level of a few percent. The experiments measure the rate of neutrino interactions,
which is the convolution of three factors: the neutrino flux, the interaction cross section and the detector efficiency. 
%Here focus on the cross sections. 
The detectors of
the modern accelerator-based neutrino oscillation experiments
are composed of complex nuclei ($^{12}$C, $^{16}$O, $^{40}$Ar, $^{56}$Fe...). 
In the hundreds-MeV to few-GeV energy region, the neutrino-nucleus cross sections
are known with a precision not exceeding 20\%, hence represent 
one of the most important sources of systematic uncertainties.
%Although a majority of interaction systematic errors can be canceled by the internal measurement of oscillation experiments mainly by the near detectors, without improving the interaction models the limitations of internal constrains remain.  
%In accelerator-based experiments the neutrino beams (at difference with respect to electron beams, for example) are not monochromatic but they span a wide range of energies. Several reaction channels can be open. 
The status of our knowledge in the different open channels in the few-GeV region, i.e. the quasielastic, the pion production and the multinucleon emission, is here briefly reviewed, devoting special emphasis to the multinucleon emission channel.
For a more detailed discussion see for example Ref.~\cite{Katori:2016yel}.

\section{CCQE, CCQE-like and CC0$\pi$}
In the discussion of the charged current quasielastic (CCQE) cross section, the MiniBooNE measurement ~\cite{AguilarArevalo:2010zc}, 
obtained using a high-statistics sample of $\nu_\mu$ events on $^{12}$C, plays a central role.
%The results were presented for the first time at the NuInt09 conference~\cite{Katori:2009du} and then published in Ref.~\cite{AguilarArevalo:2010zc}.
In this work (as well as in other experiments involving Cherenkov detectors) the quasielastic cross section is defined as the one for processes 
in which only a muon is detected in the final state, but no pions.
However it is possible that in the neutrino interaction,
a pion produced via the excitation of the $\Delta$ resonance escapes detection,
for instance because it is reabsorbed in the nucleus. 
In this case it imitates a quasielastic process. 
The MiniBooNE analysis corrected for this possibility.  %via
%%a Monte-Carlo evaluation of this process.
%a data driven correction based on the simultaneous measurement of CC1$\pi^{+}$ sample.
%The net effect amounted to a reduction of the observed quasielastic cross section. 
After applying this correction,
the quasielastic cross section thus defined still displayed an anomaly. 
The comparison of these results with a prediction based on
the relativistic Fermi gas model using in the axial form factor 
the standard value of the axial cut-off mass $M_A=1.03$ GeV$/c^2$,
consistent with the one extracted from bubble chamber experiments, reveals a substantial discrepancy. 
The introduction of more realistic theoretical nuclear models,  assuming the validity of the hypothesis that the neutrino interacts with a single nucleon in the nucleus,
does not alter this conclusion. A possible solution of this apparent puzzle was suggested %by Martini \textit{et al.}~\cite{Martini:2009uj}
in Ref.~\cite{Martini:2009uj} where the attention was drawn on the existence of additional mechanisms
beyond the interaction of the neutrino with a single nucleon in the nucleus,
which are susceptible to produce an increase of the quasielastic cross section.
The absorption of the $W$ boson by a single nucleon,
which is knocked out (the genuine quasielastic scattering),
leading to 1 particle - 1 hole (1p-1h) excitations, is only one possibility.
In addition  one must consider coupling to nucleons belonging to correlated pairs (NN correlations) and two-nucleon currents arising from meson exchange (MEC). 
This leads to the excitation of two particle -two hole (2p-2h) states.
3p-3h excitations are also possible. 
Together they are called np-nh (or multinucleon) excitations.
The addition of the np-nh excitations 
to the genuine quasielastic (1p-1h) contribution leads to an agreement with the MiniBooNE data without any increase of the axial mass.
%Isolating a genuine quasielastic event in electron scattering experiments where the kinematics (the energy and momentum transfer $\omega$ and ${\bf{q}}$) are fixed by the knowledge of the energy and momentum of incoming and outgoing electron beams is relatively easy. In the double differential cross sections, or in the nuclear responses, one can isolate the bump centered at $\omega=({\bf{q}}^2-\omega^2)/(2M_N)=Q^2/(2M_N)$ corresponding to single nucleon knockout. This is not the case in neutrino scattering experiments. Due to the broadening of the incoming neutrino flux one explores the whole energy- and momentum-transfer plane, hence the multinucleon excitations are strictly entangled with the single nucleon knockout events. This is particularly true for Cherenkov detectors, like MiniBooNE and Super-Kamiokande.This fact has fundamental consequences on the determination of the neutrino energy in the neutrino oscillation experiments. 

Processes in which only a final charged lepton is detected, hence including multinucleon excitations,
but pion absorption contribution is subtracted, today are usually called {\bf quasielastic-like}, or {\bf CCQE-like}.
Thus, what MiniBooNE published was not CCQE data, but CCQE-like data.
To avoid the confusion of the signal definition,
it is increasingly more popular to present the data in terms of the final state particles,
such as ``1 muon and 0 pion, with any number of protons''.  
This corresponds to the CCQE-like data without the subtraction of any intrinsic backgrounds
(except detector related effects) and it is called {\bf CC0$\pi$}.

After the suggestion~\cite{Martini:2009uj} of the inclusion of np-nh excitations mechanism as the likely explanation of the MiniBooNE anomaly, 
the interest of the neutrino scattering and oscillation communities on the multinucleon emission channel rapidly increased. 
Indeed this channel was not included in the generators used for the analyses of the neutrino cross sections and oscillations experiments.

\begin{figure}
\begin{center}
  \includegraphics[height=\textheight]{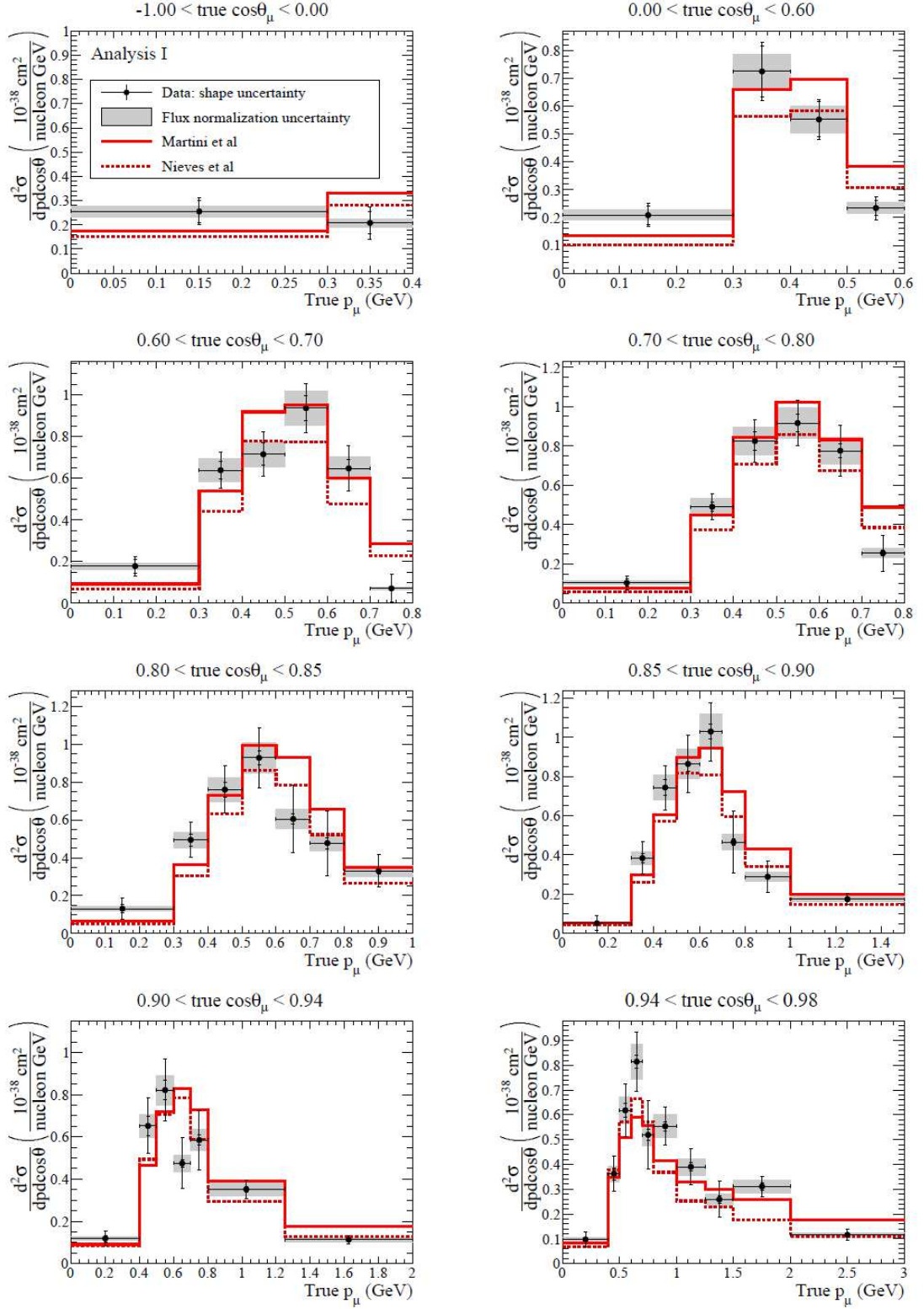}
  \caption{Double-differential muon neutrino charged-current cross section on carbon 
without pions in the final state (\textbf{CC$0\pi$}) measured by T2K using
the off-axis near detector ND280 compared with the theoretical calculations of Martini~\textit{et al.}~\cite{Martini:2009uj} and Nieves~\textit{et al.}~\cite{Nieves:2011pp}.
  The figure is taken from Ref.~\cite{Abe:2016tmq}.}
\label{fig_comp_t2k_cc0pi}
\end{center}
\end{figure}

Concerning the theoretical situation, nowadays several calculations agree on the crucial role of the multinucleon emission in order to explain the 
MiniBooNE, T2K and MINERvA cross sections. 
%flux-integrated differential cross sections which are at this moment the golden observables for the theory-experiment comparisons in neutrino scattering. 
Nevertheless there are some differences on the results obtained for this np-nh channel by the different theoretical approaches.
An illustration of the amount of the differences between the results obtained by two theoretical approaches,
is given in Fig.~\ref{fig_comp_t2k_cc0pi}, 
taken from Ref.~\cite{Abe:2016tmq}, where the \textbf{CC$0\pi$} flux-integrated double-differential cross section on carbon 
performed by T2K using the off-axis near detector ND280 is compared with the theoretical calculations of
Martini~\textit{et al.}~\cite{Martini:2009uj} and Nieves~\textit{et al.}~\cite{Nieves:2011pp}.
Flux-integrated differential cross sections in terms of the final state topology of the reaction
(as in this case CC$0\pi$ instead of CCQE or CCQE-like) are at this moment the golden observables for the theory-experiment comparisons in neutrino scattering.
%In Fig.~\ref{fig_comp_t2k_cc0pi} the T2K results are compared to the ones of Martini~\textit{et al.} and Nieves~\textit{et al.} obtained as a sum genuine quasielastic contribution calculated in RPA and np-nh excitations. As shown in Ref~\cite{Morfin:2012kn} in connection with the MiniBooNE results and in Ref.~\cite{Abe:2016tmq} for the T2K flux integrated double-differential cross sections,
As shown in Ref.~\cite{Abe:2016tmq} the two theoretical approaches give very similar results for the genuine quasielastic calculated in RPA.
The major differences are related to the np-nh channel. At the present level of experimental accuracy 
quantifying the agreement between the T2K data and the two models is not evident; the uncertainties are too large for any conclusive statement.
For the moment, from Ref.~\cite{Abe:2016tmq} one can only conclude that both models agree with the data,
and the data seems to suggest the presence of np-nh with respect to pure CCQE RPA predictions.
This is an important conclusion,
since these results represent a successful test of the necessity of the multinucleon emission channel
%in RPA based models
in order to reproduce the data of an experiment with another
neutrino flux (but in the same neutrino energy domain) with respect to the one of MiniBooNE. The detailed comparison with the MiniBooNE
$\nu_\mu$~\cite{AguilarArevalo:2010zc} and $\bar{\nu}_\mu$~\cite{AguilarArevalo:2013hm} flux integrated CCQE-like double-differential cross sections was published in Refs.~\cite{Martini:2011wp,Nieves:2011yp,Nieves:2013fr,Martini:2013sha}.  
%The same conclusion holds for the $\nu_\mu$ and $\nu_e$ CC inclusive cross sections of T2K, compared to the calculations of Martini ~\textit{et al.} in Refs. and .  
A comparison with the T2K CC$0\pi$ and MiniBooNE CCQE-like flux integrated double-differential cross sections has been recently performed also by Megias~\textit{et al.} in Ref.~\cite{Megias:2016fjk}.
Also in this case, a better agreement with data is obtained by adding the 2p-2h MEC contributions to the
genuine theoretical CCQE results calculated in the SuSAv2 approach.

In the np-nh sector the microscopic calculations of Martini~\textit{et al.}, Megias~\textit{et al.} and Nieves~\textit{et al.} are based on the Fermi gas, which is the simplest independent particle model. Even in this simple model an exact relativistic calculation is difficult for several reasons. The first difficulty is that one needs to perform 7-dimensional integrals for a huge number of 2p-2h response Feynman diagrams. Second, divergences in the NN correlations sector and in the angular distribution of the ejected nucleons 
may appear and need to be regularized. Furthermore the neutrino cross section calculations should
be performed for all the kinematics compatible with the experimental neutrino flux.
For these reasons an exact relativistic calculation is very demanding with respect to computing,
and as a consequence different approximations are employed by the different groups
in order to reduce the dimension of the integrals, and to regularize the divergences. 
The choice of subsets of diagrams and terms to be calculated also presents important differences. For a detailed discussion we refer to Ref.~\cite{Katori:2016yel}.  
Here we just mention that 
one of the major difference between the results of Megias~\textit{et al.} \cite{Megias:2014qva} and
the results of Martini~\textit{et al.} and Nieves~\textit{et al.}
related to the presence or not of 2p-2h contributions in the axial sector and in the vector-axial interference term
is now disappeared with the new results of Refs.~\cite{Simo:2016ikv,Megias:2016fjk}. 
The MEC contributions to neutrino-nucleus cross sections in the three different microscopic approaches seem now to be compatible among
them. 
The major differences that still remain in the np-nh sector, 
are related to the treatment of the NN correlations and NN correlations-MEC interference terms.
Beyond all the theoretical models mentioned above, 
other interesting calculations discussing the 2p-2h excitations in connection with the neutrino scattering recently
appeared \cite{Lovato:2014eva,Lovato:2015qka,Benhar:2015ula,Rocco:2015cil,Lovato:2016gkq,VanCuyck:2016fab}. 
For the moment no comparison with neutrino flux-integrated differential cross sections are shown however the results of these approaches,
in particular the ~\textit{ab-initio} one of Lovato ~\textit{et al.}~\cite{Lovato:2014eva,Lovato:2015qka,Lovato:2016gkq}
can offer important benchmarks for more phenomenological methods. Some examples are discussed in Ref.~\cite{Katori:2016yel}. 

The multinucleon excitations have a strong impact on the reconstruction of the neutrino energy
via the quasielastic kinematics-based method, as pointed out and discussed in several papers
~\cite{Martini:2012fa,Nieves:2012yz,Lalakulich:2012hs,Martini:2012uc,Coloma:2013tba,Ankowski:2016bji}.
This is particularly important for the determination of the neutrino oscillation parameters because data on neutrino oscillation often involve reconstructed neutrino energies while the analyses imply the real neutrino energy.
Neutrino oscillation analyses which quantitatively take into account the effect of the np-nh channel started to appear ~\cite{Abe:2015awa,Ericson:2016yjn}. 
A possible way to reduce the systematic uncertainties due to the multinucleon emission channel
in the neutrino oscillation events by maintaining
the QE-based energy reconstruction method instead of the calorimetric one
has been discussed
by Mosel~\textit{et al.} in Ref.~\cite{Mosel:2013fxa} in connection with DUNE distributions.
They suggest to consider ``CC0$\pi$1p'' sample, {\it i.e.} final state events with one charged lepton,
0 pions, and 1 proton, instead of traditional ``CC0$\pi$'' sample where the final state particles include one charged lepton and 0 pions,
and any number of protons.
%These events in CC0$\pi$1p sample are primarily due to an original QE event (about 80\% of contribution with respect to a 50\% in the CC0$\pi$ case).
This more restrictive requirement %of 0 pions and exactly 1 proton
allows to
obtain the true and reconstructed energy results quite close to each other. 
%The shift of the oscillation peaks and the tendency to concentrate the events at lower energies are drastically reduced. 
The price to pay is that one loses a factor 3 in the number of events.
Furmanski and Sobczyk proposed to include full energy-momentum conservation on CC0$\pi$1p sample to
improve the CCQE data sample and energy reconstruction~\cite{Furmanski:2016wqo}.
In order to utilize these ideas in real experiments, we need a careful evaluation of proton measurement systematics.

Precise predictions and measurements of hadronic final states are clearly the next step.
The community is moving toward to this path. CCQE-like and CC0$\pi$ cross sections of one-track (muon) and
two-track (muon and proton) samples have been published by T2K~\cite{Abe:2015oar} and MINERvA~\cite{Walton:2014esl}.
Other measurements which clearly goes in this direction are 
the one presented by MINERvA in Ref.~\cite{Rodrigues:2015hik} 
where the observed hadronic energy is combined with muon kinematics allowing to give
the results in terms of a pair of variables which separate genuine QE and $\Delta$ resonance events,
like in inclusive electron scattering experiments, and the one of ArgoNeuT on exclusive $\nu_\mu$ CC0$\pi$ events with 2 protons in the final state, the ($\mu^- + 2 p$) triple coincidence topology~\cite{Acciarri:2014gev}, like in exclusive electron scattering experiments.
From a theoretical point of view only few,
and very recent, microscopic calculations have been performed focusing on hadronic information in connection with the neutrino-nucleus scattering. We can essentially mention two studies, one of Ruiz Simo \textit{et al.} ~\cite{RuizSimo:2016ikw}
and one of Van Cuyck \textit{et al.} ~\cite{VanCuyck:2016fab}, related to the emission of nucleon pairs induced by MEC and NN short range correlations, respectively. These theoretical calculations refer to $^{12}$C. Since also other nuclear targets, 
such as $^{16}$O and $^{40}$Ar, are used in present and future neutrino experiments,
the mass dependence of multinucleon excitations, strictly related to the range of the pairs interaction, require important investigations.

%%%%%%%%%%%%%%%%%%%%%%%%%%%%%%%%%%%%%%%%%%%%%%%%%%%%%%%%%%%%%%%%%%%%%%%%%
%%
%%   use this format to include an .eps figure into your paper
%%
%\begin{figure}[htb]
%\centering
%\includegraphics[height=1.5in]{magnet}
%\caption{Plan of the magnet used in the mesmeric studies.}
%\label{fig:magnet}
%\end{figure}
%%%%%%%%%%%%%%%%%%%%%%%%%%%%%%%%%%%%%%%%%%%%%%%%%%%%%%%%%%%%%%%%%%%%%%%%%%%

\section{Pion production}
The single pion production is the largest misidentified background
for both $\nu_\mu$ - disappearance and $\nu_e$ - appearance experiments.
However, data-theory agreement remains very unsatisfactory. 
%For instance, theoretical calculations \cite{Ivanov:2012fm,Lalakulich:2012cj,Martini:2014dqa} of CC 1$\pi^+$ single and double differential cross sections as a function of muon variables are in agreement with the MiniBooNE data \cite{AguilarArevalo:2010bm}. On the contrary theoretical works \cite{Lalakulich:2012cj,Hernandez:2013jka} on the MiniBooNE differential cross sections function of the final pion variables display a reshaping of the differential cross section due to the inclusion of pion final state interaction which suppresses the agreement with the MiniBooNE data. Furthermore
Nowadays there is no model which can describe MiniBooNE \cite{AguilarArevalo:2009ww,AguilarArevalo:2010bm},
MINERvA \cite{Eberly:2014mra,McGivern:2016bwh} and T2K \cite{Abe:2016aoo} data simultaneously. %This is the so called ``pion puzzle''.
The complications of pion data analyses lay not only on their primary production models,
but also on the fact that all hadronic processes have to be modeled correctly.
Combination of data from different channels and different experiments hope to entangle and constrain all processes \cite{Katori:2016yel},
however, such an approach has been started very recently.% and currently we are struggling against this ``pion puzzle''.

\section{$\nu$ \textit{vs} $\bar{\nu}$ ; $\nu_\mu$ \textit{vs} $\nu_e$}
A precise and simultaneous knowledge of $\nu_\mu$, $\bar{\nu}_\mu$, $\nu_e$ and $\bar{\nu}_e$ cross sections
is important in connection to the oscillation experiments aiming at the
determination of the neutrino mass ordering and the search for CP violation in the lepton sector, such as T2K, NOvA, Hyper-K and DUNE.

Concerning the neutrino \textit{vs} antineutrino cross sections, it is well known that
they differ by the sign of the vector-axial interference term, the basic asymmetry which follows from the weak interaction theory.
This is the reason why the antineutrino cross sections are smaller and they falls more rapidly with the lepton scattering angle and with $Q^2$ than the neutrino ones. The presence of the vector-axial interference term introduces also an additional non-trivial asymmetry.
Due to this term the various nuclear responses weigh differently in the neutrino and antineutrino cross sections
\cite{Martini:2010ex,Ericson:2015cva}.
As a consequence the relative role of multinucleon contribution is different for neutrinos and antineutrinos.
Due to the different approximations performed by different groups to study this channel, this relative role presents some differences
in the different approaches (for a detailed discussion see Ref.~\cite{Katori:2016yel}) and represents a potential obstacle in the interpretation of experiments aimed at the measurement of the CP violation. 

Turning to the $\nu_e$ cross sections, few published experimental data exist.
This is essentially due to the relatively small component of electron-neutrino
fluxes with respect to the muon-neutrino ones hence to small statistics.
%For this reason the electron-neutrino experimental results essentially concern inclusive cross sections. 
%:Gargamelle ~\cite{Blietschau:1977mu}, T2K ~\cite{Abe:2014agb} and NOvA preliminary results ~\cite{Bu:2016grw} (a prominent exception is represented by the quasielastic measurement of MINERvA \cite{Wolcott:2015hda}).
The published flux-integrated differential cross sections are the inclusive ones of T2K ~\cite{Abe:2014agb} and the CCQE-like of MINERvA ~\cite{Wolcott:2015hda}.
The theoretical calculations of Refs.~\cite{Martini:2016eec,Megias:2016fjk,Gallmeister:2016dnq} have been compared with the T2K results ~\cite{Abe:2014agb}
and substantially agree with data. Once again this agreement needs the presence of the np-nh excitations.
The same conclusion holds also for the $\nu_e$ CCQE-like MINERvA differential cross sections on hydrocarbon \cite{Wolcott:2015hda},
compared with the SuSAv2+MEC approach in Ref.~\cite{Megias:2016fjk}. 

$\nu_\mu$ and $\nu_e$ differential cross sections have been compared in Ref.~\cite{Martini:2016eec}.
Due to the different kinematic limits, the $\nu_e$ cross sections are
in general expected to be larger than the $\nu_\mu$ ones.
However for forward scattering angles this hierarchy is opposite.
This appears for the 1p-1h excitations (genuine QE and giant resonances) at low neutrino energies.  
This behavior is related to a non-trivial dependence of momentum transfer on lepton mass and scattering angle,
and to a subtle interplay between lepton kinematic factors and response functions.
In the precision era of neutrino oscillation physics the $\nu_e$ cross sections should be known with the same accuracy as the $\nu_\mu$ ones. 
Trying to deduce the $\nu_e$ cross sections from the experimental $\nu_\mu$ ones can be considered only as
a first approximation in the study of the $\nu_e$ interactions.

\newpage
\Acknowledgements
I thank the NuPhys2016 organizers for the opportunity to present this overview.  
I am grateful to Magda Ericson for the long and continuous collaboration on neutrino-nucleus cross sections.
Finally I also thank Teppei Katori for useful discussions and for being the co-author of the review paper
\cite{Katori:2016yel} to which this overview is largely inspired.  

%\bibliography{only_marco,rev_nu_teppei}
%\bibliography{only_marco}
%\end{document}

\end{document}

%% file: econfmacros.tex
\def\beq{\begin{equation}}
\def\eeq#1{\label{#1}\end{equation}}
\def\eeqn{\end{equation}}

\def\beqa{\begin{eqnarray}}
\def\eeqa#1{\label{#1}\end{eqnarray}}
\def\eeqan{\end{eqnarray}}

\let\bar=\overbar

\def\Dslash{\not{\hbox{\kern-4pt $D$}}}
\def\dslash{\not{\hbox{\kern-2pt $\del$}}}

\def\msb{{\bar{\ssstyle M \kern -1pt S}}}